\documentclass{emulateapj}

\newcommand{\f}{\phantom{2}}

\newcommand{\umu}{\mu}

\newcommand{\al}{Al\,{\sc iii}}
\newcommand{\mg}{Mg\,{\sc ii}}

\newcommand{\civ}{C\,{\sc iv}}

\newcommand{\myemail}{chris.willott@nrc.ca}
\def\chandra{{\it Chandra~}}
\def\xmm{{\it XMM-Newton~}}

\shorttitle{The sub-millimeter properties of broad absorption line quasars}
\shortauthors{Willott et al.}

\begin{document}


\title{The sub-millimeter properties of broad absorption line quasars}


\author{Chris J.\ Willott} 
\affil{Herzberg Institute of Astrophysics, National Research Council,
5071 West Saanich Rd,\\ Victoria, B.C. V9E 2E7, Canada\\ {\tt email:
\myemail}}

\author{Steve Rawlings, Jennifer A. Grimes} 
\affil{Astrophysics, Department of Physics, Keble Road, Oxford, OX1
3RH, U.K.\\ {\tt email: sr@astro.ox.ac.uk; jag@astro.ox.ac.uk}}




\begin{abstract}

We have carried out the first systematic survey of the sub-millimeter
properties of broad absorption line (BAL) quasars. 30 BAL quasars
drawn from a homogeneously selected sample from the Sloan Digital Sky
Survey at redshifts $2<z<2.6$ were observed with the SCUBA array at
the JCMT to a typical rms sensitivity of 2.5\,mJy. Eight quasars were
detected at $>2 \sigma$ significance, four of which are at $>3 \sigma$
significance.  The far-infrared luminosities of these quasars are $ >
10^{13}\,L_{\sun}$. There is no correlation of sub-millimeter flux
with either the strength of the broad absorption feature or with
absolute magnitude in our sample. We compare the sub-millimeter flux
distribution of the BAL quasar sample with that of a sample of quasars
which do not show BAL features in their optical spectra and find that
the two are indistinguishable. BAL quasars do not have higher
sub-millimeter luminosities than non-BAL quasars. These findings are
consistent with the hypothesis that all quasars would contain a BAL if
viewed along a certain line-of-sight. The data are inconsistent with a
model in which the BAL phenomenon indicates a special evolutionary
stage which co-incides with a large dust mass in the host galaxy and a
high sub-millimeter luminosity. Our work provides constraints on
alternative evolutionary explanations of BAL quasars.

\end{abstract}




\section{Introduction}

The striking correlation between black hole mass and the mass of stars
in spheroidal galaxies suggests a physical link between the build up
of black holes by accretion and the formation of stars. Most models
for this link (e.g.\ Silk \& Rees 1998; Fabian 1999) invoke feedback
mechanisms in which black holes drive outflows: once the black hole
reaches a critical mass fixed by the host galaxy mass, these outflows
drive gas and dust out of the galaxy and terminate both star formation
and further black hole growth.

Observational evidence for outflows in quasars comes in the form of
broad absorption lines (BALs). These are metal absorption systems with
huge blue-shifted velocities (several thousand kms$^{-1}$).  The
similarity of the continuum and line emission of BAL and non-BAL
quasars motivates the hypothesis that BAL quasars are not
intrinsically different from other quasars. The presence of BAL
features in the spectra of only 15\% of optically selected quasars
(Tolea et al. 2002) could naturally be explained by a difference in
viewing angle if the sub-relativistic outflow at the origin of the BAL
features is not isotropic (Weymann et al. 1991). The popular notion
that the outflow is preferentially located close to the edge of the
torus surrounding the supermassive black hole has found support in
spectropolarimetric measurements (Goodrich \& Miller 1995; Cohen et
al. 1995). However, studies of radio-loud BAL quasars, in which the
radio properties give some information on the orientation, appear at
odds with this simple orientation model (Becker et al. 2000). The
other main contender as an interpretation of the BAL phenomenon is the
evolutionary scenario of Briggs et al. (1984) in which all quasars
pass through a BAL phase for $\sim$ 15\% of their active
lifetimes. The small sizes of the radio lobes in radio-loud BAL
quasars (Becker et al. 2000) is suggestive of the BAL phase
co-inciding with an early stage of quasar activity -- perhaps removing
a shroud of gas and dust from the nuclear region.

The sub-millimeter emission from quasars comes from optically-thin, cool dust
and is therefore orientation-independent. It is also likely heated by
young stars in starbursts which will evolve over the lifetime of the
quasar. It is therefore the prime wavelength in which to discriminate
between the orientation and evolutionary explanations of the BAL
phenomenon. If all quasars contain BALs, then BAL quasars would be
expected to have the same sub-millimeter properties as non-BAL quasars. But if
the BAL phenomenon is a phase that all quasars go through, and is
connected with the termination of large-scale star-formation, then the
sub-millimeter emission should differ between BAL and non-BAL quasars.

There are hints from previous studies with millimeter and
sub-millimeter photometry that BAL quasars are more luminous than
non-BAL quasars, but the numbers in these studies are small. Omont et
al. (1996) found that 2 out of 4 BAL quasars were detected at 1.1 mm,
whereas only 4 out of 19 non-BAL quasars were detected. More recently,
Carilli et al. (2001) observed 41 quasars from the Sloan Digital Sky
Survey (SDSS) with IRAM. However, their sample contained only 7 BAL
quasars with a detection fraction of 4/7. Compared with their non-BAL
detection fraction of 12/34, this is consistent with BALs being more
sub-millimeter luminous, but the numbers are too small to provide a
significant result. Priddey et al. (2003) detected 2 out of 8 BAL
quasars with SCUBA at the JCMT, compared to a detection rate of 7 out
of 49 non-BALs -- again suggestive of BALs being more sub-millimeter
luminous, but not at a significant level. Willott et al. (2002) and
Rawlings et al. (2003) both found that the only moderate radio-power
quasars in their samples with luminous sub-millimeter emission were
the sole objects in each sample to possess BALs.

The Sloan Digital Sky Survey (SDSS) is transforming the study of
quasars with the size, data quality and completeness of its quasar
survey. It is now possible to define very large samples of BAL quasars
with homogeneous selection criteria. BAL quasar samples have been
derived from the SDSS Early Data Release (Stoughton et al. 2002) by
Tolea et al. (2002) and Reichard et al. (2003). In this paper we
present sub-millimeter observations of a carefully selected sample of broad
absorption line quasars from the SDSS. These are compared to quasars
which do not possess BALS with similar redshift and luminosity to
determine if there is a difference between the sub-millimeter properties of
BAL and non-BAL quasars.

We assume throughout that $H_0=70\, {\rm km\,s^{-1}\,Mpc^{-1}}$,
$\Omega_{\mathrm M}=0.3$ and $\Omega_\Lambda=0.7$. All data used from
papers in the literature have been adjusted to be consistent with this
cosmological model. In Sec. 2 we describe the sub-millimeter
observations performed on the sample of BAL quasars. In Sec. 3 we
compare the sub-millimeter properties of BAL quasars with those from a
matched sample of quasars without BALs. In Sec. 4 we interpret these
results and present our conclusions in Sec. 5.

\section{Sub-millimeter photometry of BAL quasars}

\subsection{BAL quasar sample selection}

In order to determine if BAL quasars have different sub-millimeter properties
from non-BAL quasars, we require a control sample of non-BAL quasars
with matched luminosity and redshift. To avoid doubling our required
observing time by defining our own control sample, we instead defined
our BAL sample so that it is similar to non-BAL sub-millimeter/millimeter quasar
surveys performed by others. Since BAL quasars are usually defined on
the basis of absorption of the \civ\ or \mg\ lines, highly complete
samples are limited to $z<4$ (e.g. Reichard et al. 2003). Further,
since the SDSS contains many more quasars at $z\sim2$ than $z\sim3$,
to define a large sample of BAL quasars we need to work at redshifts
$\sim2$. Two recent quasar surveys have been carried out at this
redshift with SCUBA at the JCMT (Priddey et al. 2003) and MAMBO at the
IRAM 30-m (Omont et al. 2003) and we combine these for the control
sample (further details are given in Sec.\, 3.1).

To select a sample of BAL quasars for observation we use the SDSS EDR
BAL sample of Reichard et al. (2003), in preference to that of Tolea
et al. (2002), since it utilized a more detailed selection method. Our
BAL quasar sample is restricted to the redshift range $2<z<2.6$ to be
comparable with the bulk of sources in the control sample. We further
restrict the sample to the RA range 10$^{\rm h}$ - 18$^{\rm h}$ for
compatibility with telescope scheduling. The balnicity index BI gives
a measure of the strength of broad absorption features (e.g. Weymann
et al. 1991). All the quasars in our redshift range show the \civ\
wavelength region in their spectra, so our selection is on the basis
of the \civ\ balnicity. The Reichard et al. sample contains 39 quasars
in this redshift and RA range with a wide range of BI
($0-10\,000\,{\rm km\,s}^{-1}$). The four quasars with BI~$<200\,{\rm
km\, s}^{-1}$ were excluded from our sample since these represent very
weak broad absorption features. Of the remaining 35 quasars we
selected a sample of 30 with the highest dereddened absolute
magnitudes. We made a cut in absolute magnitude because we did not
have time to observe all 35 sources and by just selecting the most
luminous we enable the luminosity distribution to be more similar to
that of the control sample. Reichard et al. (2003) found that the
optical colors of BAL quasars are redder than those of non-BAL
quasars. This can naturally be explained by dust-reddening. A
comparison of composite quasar spectra formed by matched BAL and
non-BAL SDSS sub-samples shows the mean amount of reddening is
$E(B-V)=0.023$. We determine the dereddened absolute $B$-band
magnitudes of the BAL quasars assuming a rest-frame color of
$B-i^{\star}=+0.35$ (Schneider et al. 2002) and incorporating an
offset of $-0.2$ magnitudes to account for reddening. The absolute
magnitude cut to restrict the sample to 30 quasars is at
$M_B<-26.6$. Only three of these 30 BAL quasars are detected in the
radio. These three quasars have 1.4\,GHz flux-densities in the range
2-10\,mJy, so the contribution due to synchrotron emission at 850\,$\umu$m
is expected to be much less than 1\,mJy and hence can be neglected (Willott
et al. 2002).

\begin{deluxetable*}{lccccrr}
\tabletypesize{\scriptsize}
\tablecaption{Table of observed objects. \label{tbl-1}}
\tablewidth{0pt}
\tablehead{
\colhead{Source} &     \colhead{$z$}   & \colhead{$M_B$} & \colhead{BI$~ /~ {\rm km\,s}^{-1}$}  & \colhead{Type} & \colhead{$S_{850}~ /~  $mJy} & \colhead{$S_{450}~ /~  $mJy} 
}
\startdata
SDSS\,J104109.85$+$001051.8  &    2.250   &     $-26.68$  &     1913 \f  &      Hi      &      $   0.54   \pm 2.26$ \f  &                     \f   \\
SDSS\,J104130.17$+$000118.8  &    2.068   &     $-27.66$  &  \f  672 \f  &   \f   Lo?   &      $   2.04   \pm 2.73$ \f  &                     \f   \\
SDSS\,J104233.86$+$010206.3  &    2.123   &     $-27.20$  &  \f  401 \f  &      Hi      &      $  -0.56   \pm 2.80$ \f  &                     \f   \\
SDSS\,J104841.02$+$000042.8  &    2.022   &     $-26.74$  &     1176 \f  &      Hi      &  $ {\bf  4.22}  \pm 2.00$ \f  &                     \f   \\
SDSS\,J110041.19$+$003631.9  &    2.017   &     $-27.30$  &     4687 \f  &      Hi      &      $   0.03   \pm 2.13$ \f  &                     \f   \\
SDSS\,J110623.52$-$004326.0  &    2.450   &     $-27.80$  &     4034 \f  &   \f   Lo?   &      $  -0.93   \pm 3.26$ \f  &                     \f   \\
SDSS\,J121633.90$+$010732.8  &    2.018   &     $-27.04$  &     2200 \f  &      Hi      &      $   3.45   \pm 2.62$ \f  &  $   26.7 \pm 14.9$ \f   \\
SDSS\,J121803.28$+$001236.8  &    2.010   &     $-26.90$  &  \f  269 \f  &      Hi      &  $ {\bf  5.44}  \pm 2.47$ \f  &                     \f  \\ 
SDSS\,J122228.39$-$011011.0  &    2.284   &     $-26.70$  &  \f  678 \f  &   \f   Lo?   &      $  -0.32   \pm 2.15$ \f  &                     \f  \\ 
SDSS\,J123056.58$-$005306.3  &    2.162   &     $-26.91$  &  \f  652 \f  &      Hi      &      $   1.29   \pm 3.05$ \f  &                     \f  \\ 
SDSS\,J124720.27$-$011343.1  &    2.283   &     $-26.81$  &  \f  336 \f  &      Hi      &  $ {\bf  6.88}  \pm 1.91$ \f  &                     \f  \\ 
SDSS\,J131853.45$+$002211.4  &    2.079   &     $-26.97$  &     1473 \f  &      Hi      &      $   4.45   \pm 3.08$ \f  &                     \f  \\ 
SDSS\,J134145.13$-$003631.0  &    2.205   &     $-27.17$  &  \f  870 \f  &    FeLo\f\f  &      $   0.17   \pm 2.66$ \f  &                     \f  \\ 
SDSS\,J134544.55$+$002810.8  &    2.516   &     $-27.46$  &     1510 \f  &      Hi      &      $  -3.14   \pm 2.54$ \f  &  $    1.1 \pm 11.8$ \f   \\
SDSS\,J135317.80$-$000501.3  &    2.320   &     $-26.73$  &     9821 \f  &      Lo      &      $   4.17   \pm 2.73$ \f  &                     \f  \\ 
SDSS\,J135559.04$-$002413.6  &    2.332   &     $-27.65$  &     1525 \f  &      Hi      &  $ {\bf  4.15}  \pm 1.94$ \f  &                     \f  \\ 
SDSS\,J135721.77$+$005501.1  &    2.001   &     $-27.48$  & \f   235 \f  &   \f   Hi?   &      $   4.14   \pm 2.60$ \f  &  $   39.5 \pm 21.8$ \f   \\
SDSS\,J140918.72$+$004824.3  &    2.000   &     $-27.02$  &     1411 \f  &      Hi      &  $ {\bf  9.99}  \pm 3.01$ \f  &  $   23.1 \pm 15.7$ \f   \\
SDSS\,J142050.34$-$002553.1  &    2.103   &     $-26.95$  &     3442 \f  &   \f   Lo?   &      $   3.47   \pm 2.33$ \f  &  $ -26.9  \pm 25.8$ \f   \\
SDSS\,J143022.47$-$002045.2  &    2.544   &     $-26.61$  &     1957 \f  &      Hi      &      $  -1.62   \pm 2.78$ \f  &                     \f  \\ 
SDSS\,J144256.86$-$004501.0  &    2.226   &     $-27.51$  & \f   815 \f  &      Hi      &      $   4.33   \pm 2.35$ \f  &  $    9.6 \pm 21.7$ \f   \\
SDSS\,J145045.42$-$004400.3  &    2.078   &     $-27.27$  & \f   238 \f  &      Hi      &      $   1.85   \pm 2.38$ \f  &  $ -31.8  \pm 23.1$ \f   \\
SDSS\,J150033.52$+$003353.7  &    2.451   &     $-27.86$  &     4257 \f  &      Hi      &  $ {\bf  8.13}  \pm 2.56$ \f  &                     \f  \\ 
SDSS\,J150206.66$-$003606.9  &    2.202   &     $-27.17$  & \f   406 \f  &      Hi      &      $   0.29   \pm 2.51$ \f  &  $   11.9 \pm 22.2$ \f   \\
SDSS\,J151636.79$+$002940.4  &    2.240   &     $-28.44$  &     4035 \f  &      Lo      &      $  -1.37   \pm 2.50$ \f  &  $    3.0 \pm 15.0$ \f   \\
SDSS\,J152913.85$-$001013.8  &    2.073   &     $-27.31$  & \f   209 \f  &      Hi      &      $  -1.31   \pm 2.52$ \f  &  $   30.5 \pm 22.4$ \f   \\
SDSS\,J170056.85$+$602639.8  &    2.125   &     $-27.08$  &     1400 \f  &      Hi      &  $ {\bf  6.62}  \pm 2.64$ \f  &                     \f  \\ 
SDSS\,J171652.35$+$590200.2  &    2.369   &     $-27.16$  &  \f  620 \f  &      Hi      &  $ {\bf 10.53}  \pm 3.04$ \f  &                     \f  \\ 
SDSS\,J172012.40$+$545601.0  &    2.099   &     $-27.33$  &     1249 \f  &      Hi      &      $   3.75   \pm 2.06$ \f  &  $   27.9 \pm 21.9$ \f   \\
SDSS\,J172341.09$+$555340.5  &    2.113   &     $-27.11$  &     3497 \f  &    FeLo\f\f  &      $  -3.74   \pm 2.28$ \f  &  $ -10.4  \pm 17.6$ \f   \\
 \enddata


\tablecomments{SCUBA photometry of SDSS broad absorption line quasars
at $2<z<2.6$ drawn from the sample of Reichard et al. (2003). The
$B$-band absolute magnitudes are derived from the SDSS EDR
$i^\star$-band absolute magnitudes by (i) a correction to our adopted
cosmology; (ii) a color correction of $B-i^\star=+0.35$ (Schneider et
al. 2002); (iii) dereddening by $-0.2$ magnitudes to account for the
average amount of intrinsic dust reddening compared to non-BAL
quasars. Balnicity indices BI are for the \civ\ line and were computed
by fitting a composite quasar spectrum to the continuum (column 12 of
Table 1 of Reichard et al. 2003). Type denotes whether the quasar is a
Hi-ionization, Lo-ionization or Fe-Lo-ionization BAL (column 18 of
Table 1 of Reichard et al. 2003 with `?' from column 17 of the same
table). 850\,$\umu$m detections at the $>2 \sigma$ level are
highlighted in a bold font. The three quasars detected in the radio are SDSS\,J135721.77$+$005501.1, SDSS\,J150206.66$-$003606.9 and SDSS\,J151636.79$+$002940.4. None of them are detected at $>2 \sigma$ at 850\,$\umu$m. 450\,$\umu$m flux-densities are only given
for those observations performed in good conditions where the data for
the 450\,$\umu$m array is reliable. There are no $>2 \sigma$
detections at 450\,$\umu$m. }
\end{deluxetable*}

\subsection{Observations}

The entire sample of 30 BAL quasars was observed in photometry mode
with the SCUBA bolometer array at the JCMT from February to June
2003. Observations were made simultaneously at 850\,$\umu$m and 450
$\umu$m. The majority of the observations were performed in good to
average conditions (atmospheric zenith opacity at 1300\,$\umu$m in the
range $0.05-0.12$) and hence the 450\,$\umu$m data are not nearly as
effective at detecting thermal dust emission as the 850\,$\umu$m
data. Henceforth we will concentrate on the 850\,$\umu$m data, but the
450\,$\umu$m data from some observations are also given in Table 1.
The source was chopped between two bolometers on the array to increase
the effective time on source.

The data were reduced using the SURF package, following mostly the
same methods as in Archibald et al. (2001). The main difference is
that the sky opacity was determined from polynomial fits to either the
JCMT water vapor monitor or, on the occasions when this was not
operating, the NRAO 350\,$\umu$m tau meter. The data for the two
bolometers were reduced separately and the final fluxes measured with
each bolometer were combined weighting by the inverse of the
variance. For some observations the secondary bolometer data are
unstable and for these only data from the primary bolometer is used.

The resulting flux-densities are given in Table 1. We have achieved
our aim of reaching an rms sensitivity at 850\,$\umu$m lower than 3.33
mJy (i.e. $3\sigma=10$\,mJy) for all of the 30 broad absorption line
quasars observed. In Table 1 we highlight in bold the eight quasars
which are detected at the $>2 \sigma$ level. Of these, four are
securely detected at the $>3 \sigma$ level.  A quick glance at the
table shows that the detected quasars span a range of $z$, $M_B$ and
balnicity index. A full discussion of the sample statistics and a
comparison with the control sample is pursued in the following
section. None of the quasars are detected at even the $>2 \sigma$
level at the shorter wavelength of 450\,$\umu$m. This is not
surprising given the sensitivity level reached. These data do not
provide useful constraints on the sub-millimeter spectral indices.

We estimate the submillimeter/far-infrared luminosity from the
sub-millimeter fluxes by assuming the isothermal dust template of
Priddey \& McMahon (2001).  This template consists of isothermal
optically-thin dust with temperature $T=41$ K and emissivity index
$\beta=1.95$. This gives a conversion between $L_{\rm FIR}$ and
$S_{850}$ of $L_{\rm FIR}\approx2\,\times\,10^{12}\, (S_{850}\, /\,
{\rm mJy})\, L_{\sun}$ at redshifts $z>1$.  However, as discussed in
Willott et al. (2002) these values of $L_{\rm FIR}$ should be treated
as lower limits since isothermal models likely underestimate the total
far-infrared luminosity. The four BAL quasars we have detected with
SCUBA at $> 3 \sigma$ significance have far-infrared luminosities $ >
10^{13}\,L_{\sun}$. In the rest of this paper we will use the observed
submillimeter flux-density in preference to the far-infrared
luminosity and thereby avoid uncertainty in $L_{\rm FIR}$ due to the
unknown dust spectral shape.

\section{Do quasars with BALS differ from those without?}

\subsection{Control sample}

In this section we combine our new data with data from the literature
to determine how the sub-millimeter emission depends upon whether a
quasar shows broad absorption lines or not. As mentioned previously,
our sample of BAL quasars was defined so as to be as similar as
possible in terms of redshift and luminosity as quasars already
observed in the sub-millimeter or millimeter. Priddey et al. (2003)
obtained 850\,$\umu$m data for 57 quasars in the redshift range
$1.5<z<3$ with SCUBA on the JCMT. Their sample is drawn from the Large
Bright Quasar Survey (LBQS; Hewett, Foltz \& Chaffee 1995) and Hamburg
Quasar Survey (HS; Hagen et al. 1999). Omont et al. (2003) obtained
1.2\,mm data for 35 quasars in the redshift range $1.8<z<2.8$ with
MAMBO at the IRAM 30-m telescope. Their sample was drawn from the
V\'eron-Cetty \& V\'eron (2001) catalog and has some overlap with the
HS quasars observed by Priddey et al. The 1.2\,mm fluxes and errors
from Omont et al. were multiplied by a factor of 2.5 to estimate the
corresponding values at 850\,$\umu$m (Omont et al. 2003; see also
Eales et al. 2003). In cases where both papers present data on the
same source, we take the data from Priddey et al. in favor of that of
Omont et al., since these observations were made at the same
wavelength as our data and do not require multiplication by a flux
correction factor. Note that there are a couple of cases of detections
by Omont et al. and non-detections by Priddey et al.  However, to take
the Omont et al. data for these sources would bias the sample
unfairly. Both these papers contained a few sources showing BALs which
we exclude, since we require a non-BAL control sample. Following
Priddey et al. (2003) and matching to the sensitivity reached for the
BAL sample, we exclude from further analysis the three sources from
the Priddey et al. sample and the two sources from the Omont et
al. sample for which the observations were noisy (rms at 850\,$\umu$m
$>3.33$\,mJy). In summary, our control sample at $2<z<2.6$ contains 25
quasars from Priddey et al. and 16 quasars from Omont et al.

Unfortunately for us, both the above surveys concentrated on the most
optically-luminous quasars to enable comparison with high-luminosity
quasars at $z\approx4$. Our sample is drawn from the SDSS which has
fainter limiting magnitudes than the above surveys and hence our
sample contains typically less-luminous quasars. There has been no
systematic sub-millimeter/millimeter survey of $z\approx 2$ SDSS
quasars which would have provided the ideal control sample to ours.
The median $B$-band absolute magnitudes for the samples are
$M_B=-27.2$ for the BAL sample and $M_B=-28.2$ for the non-BAL
sample. Applying the Kolmogorov-Smirnov statistic to the two samples,
we find that their distributions in $M_B$ differ very significantly
(probability $10^{-8}$\%). We can also use the Kolmogorov-Smirnov statistic to
determine the offset in $M_B$ to be applied to the BAL quasar sample
to make its $M_B$ not signifcantly different from the non-BAL sample
at the 5\% level. We find that a shift of $\delta M_B =-0.8$ is needed
to make the two samples have similar flux distributions.

Omont et al. (2003) combined several different samples to
investigate whether there is a correlation between the optical and
sub-millimeter luminosities of quasars and found no statistically
significant correlation between the two luminosities. However, they
also noted that at $z\approx 4$ the weighted mean sub-millimeter flux
of an SDSS-based sample (Carilli et al. 2001) was half that of the
weighted mean sub-millimeter flux of a more luminous sample (median
$M_B$ brighter by 1 magnitude; Omont et al. 2001). This is suggestive
that there may be a weak correlation between the optical and
sub-millimeter luminosities that is shrouded by the large scatter. We
will return to this issue later.

We have also checked to see if the two samples are matched in terms of
redshift. The redshift ranges for the two samples are identical with
$2.0<z<2.6$, however the BAL sample has a median $z=2.16$ and the
non-BAL sample a median $z=2.27$. Applying the Kolmogorov-Smirnov
statistic to the two samples, we find that the probability that the
two samples are drawn from the same distribution in redshift is
4\%. There is evidence here that the samples are not ideally matched
in redshift, but the difference in cosmic time between the median
redshifts is minimal and there can not have been significant evolution
of the universe over this period, so this will not affect our results.

\begin{figure}
\resizebox{0.48\textwidth}{!}{\includegraphics{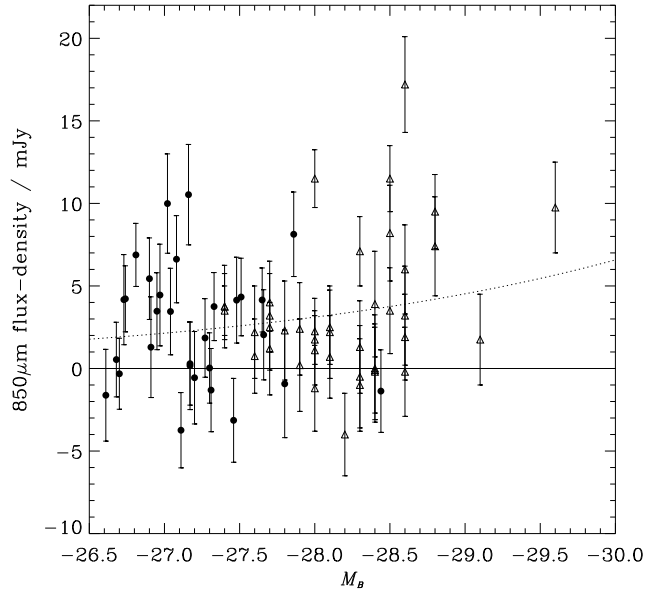}}
\caption{850\,$\umu$m flux-density against absolute magnitude for the BAL
quasar sample (filled circles) and the non-BAL control sample (open
triangles). Error bars show the $1 \sigma$ uncertainty on the measured
fluxes. A comparison between the two samples and possible correlations
in this plot are discussed in Sec.\,3. The dotted line shows a
positive correlation between sub-millimeter flux and optical
luminosity where a difference in $M_B$ of 1 equals a change in
$S_{850}$ of a factor of $1.5$.}
\label{fig:mb}
\end{figure}

\begin{deluxetable*}{lrccccccc}
\tabletypesize{\scriptsize}
\tablecaption{Statistics of quasar samples at $2<z<2.6$ surveyed in the (sub)-millimeter. \label{tbl-2}}
\tablewidth{0pt}
\tablehead{
\colhead{Reference} &    \colhead{QSO type} &    \colhead{$N$}   & \colhead{median rms} & \colhead{$<S_{850}>$}  & \colhead{Weighted $<S_{850}>$} & \colhead{median $S_{850}$} &  \colhead{$\% (S_{850}>10$)}  &  \colhead{$\% (S_{850}>6.66$)} 
 }
\startdata
This paper &     BAL & 30 & 2.54 & $2.56 \pm 0.67$  &  $2.55 \pm 0.45$  &  3.5  &  \f3 &  13 \\ 
P+O        & Non-BAL & 41 & 2.70 & $3.34 \pm 0.64$  &  $3.61 \pm 0.37$  &  2.3  &  \f7 &  20 \\ 
P          & Non-BAL & 25 & 2.80 & $2.88 \pm 0.79$  &  $2.83 \pm 0.55$  &  2.2  &  \f4 &  16 \\ 
O          & Non-BAL & 16 & 2.25 & $4.05 \pm 1.09$  &  $4.29 \pm 0.51$  &  3.5  &  13  &  25 \\ 
\enddata


\tablecomments{Sub-millimeter properties of various samples of quasars
at redshift $2<z<2.6$. All fluxes in this table, $S_{850}$, are at
850\,$\umu$m and are in units of mJy. References: P -- Priddey et
al. (2003); O -- Omont et al. (2003); the subsets of the data in these
papers used here is described in Sec.\,3.1. $N$ is the number of the
sources in the sample used for the statistical analyses in this paper.
$<S_{850}>$ is the mean observed sub-mm flux of the sample (not
weighted by the noise) and the associated standard error. Weighted
$<S_{850}>$ is the weighted mean sub-mm flux calculated as
$\Sigma(w_iS_i)/\Sigma(w_i)$ where the weights are $w = 1/\sigma^2$
and the uncertainty in $<S_{850}>$ is $[\Sigma(w_i)]^{-0.5}$.  }
\end{deluxetable*}

\subsection{Statistical analysis}

The main problem we are interested in is whether there is a difference
in the sub-millimeter properties of BAL and non-BAL quasars. We have
used the observed 850\,$\umu$m fluxes to calculate the mean, standard
error on the mean and median fluxes for four different samples: (i)
the BAL quasars of this paper; (ii) the combined Priddey et al. and
Omont et al. sample of non-BALs in the same redshift range; (iii) only
the Priddey et al. non-BALs; (iv) only the Omont et al. non-BALs. We
have also used the observed 850\,$\umu$m fluxes with their statistical
uncertainties to calculate the weighted mean and its associated
error. These results are presented in Table 2. The median sensitivity
reached in our survey is similar to that in the non-BAL control sample
(P+O). Both the mean fluxes and the weighted mean fluxes of the two
samples are comparable within the uncertainties. The non-BAL sample
actually shows slightly higher mean fluxes although the BAL sample has
a higher median flux. Also shown in the table are the percentages of
sources which are detected at $>10$\,mJy and $>6.66$\,mJy. We do not
consider the percentage of $>3 \sigma$ and $>2 \sigma$ detections
since these numbers depend critically on the sensitivity distributions
of the samples. We find that the BAL sample contains a lower
percentage of these sub-mm luminous sources than the non-BAL
sample. However, the percentage is similar to that seen in the Priddey
et al. sub-sample. 

Due to the large number of 850\,$\umu$m upper limits in both samples,
we also use survival analysis statistical tests which can account for
these limits (Feigelson \& Nelson 1985; Isobe, Feigelson \& Nelson
1986). To minimise the number of upper limits, we consider sources
with snr$>2$ to be detections for these tests. We wish to determine
whether the 850\,$\umu$m flux distributions of the BAL and non-BAL
quasars are different. Using a variety of tests (the Gehan, logrank
and Peto-Prentice tests), the returned probabilities that the fluxes
of the BAL and non-BAL samples are drawn from the same distribution
lie in the range (54-99\%). Again, we find no evidence that BAL and
non-BAL quasars differ in their sub-millimeter fluxes.

\subsection{Effects of a correlation between sub-millimeter and 
optical luminosities}

In the previous section we performed several different statistical
tests and found with each one that there is no difference in the
sub-millimeter properties of our samples of BAL and non-BAL
quasars. As mentioned in Sec.\,3.1, the fact that the luminosity
distributions of the two samples are not exactly matched leads to the
uncertainty that a positive correlation between optical and
sub-millimeter luminosity combined with enhanced sub-millimeter
luminosity in BAL quasars could have led to the observed results. We
now consider whether a correlation between the optical and
sub-millimeter luminosities exists and, if it does, whether it could
have affected our results.

We use the survival analysis bivariate correlation and regression
tests (Isobe, Feigelson \& Nelson 1986) to assess the signficance of
any correlations and their slope in these datasets. For the sample of
30 BAL quasars, the probability that there is no correlation between
850\,$\umu$m flux and $M_B$ is 45\% for Cox's proportional hazard
model and 72\% for the generalized Kendall's tau test. The
best-fitting slope for the regression of $\log_{10} S_{850}$ and $M_B$
is $0.03 \pm 0.12$. Therefore, there is no correlation present between
these quantities for the BAL sample.

We now check for a correlation between these same variables in the
non-BAL sample formed from the combination of the Priddey et
al. (2003) and Omont et al. (2003) data. We now find a very
significant (probabilities of 0.05\% and 0.3\% using the two tests)
anti-correlation between these quantities, i.e. a positive correlation
between the sub-millimeter and optical luminosities. This is quite
surprising since Priddey et al. and Omont et al.  performed similar
statistical tests on each of their full samples and did not find any
significant correlation. The reason that we have now detected this
correlation may be that by combining the samples over a restricted
redshift range we have disentangled luminosity and redshift effects.

It is important to remember that some of the most luminous observed
quasars at these redshifts are gravitationally lensed and if a
substantial number of these quasars are magnified by even modest
factors of $\approx 2$, then this would induce a correlation between
$S_{850}$ and $M_B$ even if one does not exist intrinsically. We do
not know which of these quasars are gravitationally lensed, but repeat
the correlation tests for subsets of the data in an attempt to
minimise the effects of lensing. First, we repeat the tests excluding
the two outliers on Fig. \ref{fig:mb} -- the source with
$S_{850}=17.2$\,mJy and the one with $M_B=-29.6$. We still find
significant probabilities of 0.9\% and 2\% that there is no
correlation present. Linear regression with the parametric EM
algorithm gives the slope of the correlation between $\log_{10}
S_{850}$ and $M_B$ as $-0.23 \pm 0.10$. This means that a difference
in $M_B$ of 1 equals a change in $S_{850}$ of a factor of $1.7 \pm
0.4$. We now repeat the tests using just the 30 non-BAL quasars with
the faintest absolute magnitudes, since if lensing is an issue it is
most likely to affect the quasars with the brightest observed absolute
magnitudes. The correlation probabilities are now 10\% and 14\% for
the two tests - suggestive of a correlation but not significant at the
5\% level. The slope of the correlation between $\log_{10} S_{850}$
and $M_B$ is $-0.16 \pm 0.17$, i.e. a difference in $M_B$ of 1 equals
a change in $S_{850}$ of a factor of $1.5 \pm 0.5$. This correlation
is shown as a dotted line in Fig. \ref{fig:mb}.

We conclude that whilst it is highly likely that a correlation between
$S_{850}$ and $M_B$ does exist at some level, the scatter and weakness
of the correlation mean that it is hard to prove significantly with
current data for optically-selected quasars. Similar conclusions were
reached by Omont et al. and we refer the reader to their discussion on
the likely causes of such a correlation. Grimes et al. (2003) have
investigated the $S_{850}$-$M_B$ correlation for radio-loud quasars,
which assuming that $S_{850}$ is not strongly influenced by whether or
not a quasar has radio jets, would be indicative of the relation for
quasars in general. They find that $S_{850}$ drops by a factor of 1.5
per unit drop in $M_B$. This kind of correlation is similar to that
found in the non-BAL sample.

At the low-luminosity end of the quasar luminosity function, sources
are generally too weak to be detected individually with an instrument
such as SCUBA. Stacking analyses of faint, soft X-ray sources from
\chandra and \xmm show a positive mean flux. Combining the results of
Barger et al. (2001) and Waskett et al. (2003) gives a mean
sub-millimeter flux for faint soft X-ray sources of $S_{850} =0.60 \pm
0.15$\,mJy. The optically-unobscured AGN in these samples have
$M_{B} \approx -21$. This is 7 magnitudes fainter than the non-BAL
sample which has a mean sub-millimeter flux of $S_{850} =3.34 \pm
0.64$\,mJy. The correlation between $S_{850}$ and $M_B$ implied by
these two points is that $S_{850}$ drops by a factor of 1.3 per unit
drop in $M_B$. This is somewhat less strong than the factor of 1.5 per
unit drop in $M_B$ found at the brighter end of the luminosity
function, but is further evidence that the correlation is likely not
steeper than this.

For the purposes of our BAL quasar survey we wish to know if the
existence of a correlation between $S_{850}$ and $M_B$ could affect
our findings. The median absolute magnitude of the BAL sample is
$M_B=-27.2$ and for the non-BAL sample it is $M_B=-28.2$. Therefore,
we scale the observed sub-millimeter fluxes of the BAL sample by a
factor of 1.5 (except for those with negative flux) to simulate the
effects of a correlation with slope similar to that observed in the
non-BAL sample. We repeated the two sample tests with these scaled BAL
fluxes and the comparison sample of non-BAL quasars. The returned
probabilities that the sub-millimeter fluxes of the BAL and non-BAL
samples are drawn from the same distribution are 14\% and 27\%.  This
test shows that, even after allowing for the effects of a weak
correlation between sub-mm flux and $M_B$, there is not a significant
difference between BAL and non-BAL quasars.

\subsection{Correlation with balnicity}

\begin{figure}
\resizebox{0.48\textwidth}{!}{\includegraphics{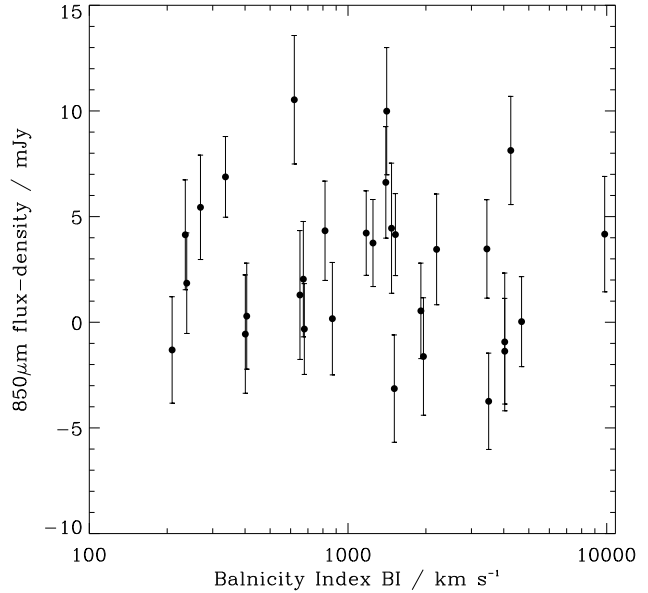}}
\caption{850\,$\umu$m flux-density against balnicity index BI
for the BAL quasar sample. There is no correlation present between BI
and sub-millimeter flux (details are given in Sec. 3.4).}
\label{fig:bi}
\end{figure}

The new sample of SDSS BAL quasars we are using was selected using
methods that are very sensitive to the presence of weak BALs (Reichard
et al. 2003). Quasars with extremely weak BAL features (BI~$<200\,{\rm
km\, s}^{-1}$) were excluded from our sample, but 43\% of our sample
is made up of quasars with moderate balnicities of
$200<$~BI~$<1000\,{\rm km\, s}^{-1}$. It is therefore important to
check whether our results could be affected by a correlation between
sub-millimeter emission and the balnicity index. In Fig. \ref{fig:bi}
we plot the 850\,$\umu$m flux against the balnicity index for our
sample. The control sample is not shown since these are non-BAL
quasars with BI~$=0\,{\rm km\, s}^{-1}$ and lie to the left of the
plotted region.  It is clear that there is no strong correlation
between 850\,$\umu$m flux and BI and both the four $>3\sigma$
detections and the sources with negative measured fluxes are spread
throughout the range of BI. Using the survival analysis correlation
routines we find that both the Cox and Kendall's Tau tests show there
is no correlation between BI and 850\,$\umu$m flux (the two tests give
54\% and 72\% probabilities for no correlation present,
respectively). One can also divide the two samples at the median
balnicity of BI~$=1400\,{\rm km\, s}^{-1}$ and calculate the weighted
means of the low- and high-BI sub-samples as $2.99 \pm 0.62$\,mJy and
$2.07 \pm 0.65$\,mJy, respectively. Once again, there is no indication
of a correlation between BI and 850\,$\umu$m flux.

BAL quasars are usually classified into Hi- and Lo-ionization BALs,
dependent upon whether low-ionization broad absorption lines such as
\mg\ and \al\ are present. LoBALs are more prevalent among
infrared-selected samples (Boroson \& Meyers 1992) and are subject to
greater dust reddening than HiBALs (Reichard et al. 2003), implying
they could be enshrouded in a dusty cocoon during an early
evolutionary stage of quasar activity (Voit, Weymann \& Korista
1993). It is therefore interesting to see if the sub-millimeter
properties are correlated with the classification. In Table\,1 we give
the BAL classifications for these quasars from Reichard et al. There
are 22 HiBALs, 6 LoBALs (of which 4 have uncertain low-ionization
troughs and definite high-ionization troughs) and 2 FeLoBALs (see Hall
et al. 2002). All of the eight $>2\sigma$ detections are in the 22
HiBALs. This is the opposite result to that expected if LoBALs reside
in more dusty environments. However, given the uncertainty surrounding
the classification of some of the LoBALs and the fact that \mg\ moves
out of the SDSS spectral range at $z>2.26$ means that these
classifications are not too reliable and no firm conclusions can be
drawn. It is certainly possible that if a sample of LoBALs were
observed in the future with SCUBA that they could have a different
distribution of sub-millimeter flux than non-BAL quasars.

\section{Interpretation}

The analyses presented in Sec.\,3 have clearly shown that quasars with
broad absorption lines show the same sub-millimeter properties as quasars
without broad absorption lines. What does this tell us about the
nature of the BALs and the evolution of quasars? The first question we
are attempting to answer is ``are BAL quasars linked to non-BAL
quasars by an evolutionary sequence or by orientation?''. Since sub-millimeter
emission is optically-thin there should be no orientation dependence
of sub-millimeter emission. Our finding that the sub-millimeter emission is not
related to the presence of a BAL is therefore consistent with the
orientation hypothesis that all quasars contains BALs, but only a
fraction of them are visible along our line-of-sight.

The evolutionary hypothesis is slightly more complicated since there
are no reliable models for the relative timings of star-formation and
AGN activity and the associated sub-millimeter emission and BAL formation.  In
the evolutionary scenario, broad absorption lines are observed at a
special epoch in quasar evolution when two conditions are met: (i) the
quasar luminosity becomes high enough to produce powerful winds that
are capable of accelerating gas to several thousand km\,s$^{-1}$; (ii)
the nuclear regions contain a considerable amount of diffuse gas and
dust. The black holes and hence maximum attainable luminosities of
quasars grow with time as matter is accreted.  Only when the
luminosity reaches a certain value will the quasar be capable of
producing strong winds. One can imagine a scenario where there is a
brief period after this luminosity has been attained in which the bulk
of the diffuse gas is ejected from the galaxy and the object shows
broad absorption lines along most lines-of-sight (see Fabian 1999 for
rough calculations). After this, the quasar will continue to shine for
as long as it has fuel (which presumably is not ejected because it is
in a denser accretion disc geometry) but no longer shows broad
absorption lines.

How does this picture relate to the expected sub-millimeter emission?
Sub-millimeter emission comes from cool dust ($\sim 50$\,K) which is
heated either by hot young stars or the active nucleus. The heating
source for quasars with luminous sub-millimeter emission is a
long-disputed issue (e.g. Rowan-Robinson 2000) but the mere presence
of a large dust mass implies that star-formation did not cease a long
time ago (Hughes, Dunlop \& Rawlings 1997).  Furthermore, the
existence of large molecular gas masses in {\em some}
sub-millimeter-luminous quasars (e.g. Combes 2001) shows that these
host galaxies are at early stages in their evolution. For our purposes
it is enough to know that diffuse, widely distributed dust close to the
nucleus and/or star-forming regions will produce large amounts of
sub-millimeter emission. The gas responsible for BALs also
contains dust (the redder optical colours of BALs compared to non-BAL
quasars is evidence of this; Reichard et al. 2003; Richards et
al. 2003). In the evolutionary scenario, the dust gets driven away
from the galaxy along with the BAL gas and therefore one would expect
that BAL quasars should show considerably stronger sub-millimeter
emission than non-BAL quasars.  Our data do not support this simple
model and show that the dust mass heated by star-formation and/or the
AGN is similar for BALs and non-BALs.

This does not mean that our results disprove the evolutionary scenario
and support the unification by orientation of BAL and non-BAL quasars,
merely that the particular picture outlined above is ruled out. One
possible modification is that the bulk of the dust which produces a
sub-millimeter luminous phase is not co-incident (in space and/or time) with
the BAL gas which is ejected. For example, Archibald et al. (2002)
present a model whereby the bulk of the star-formation and sub-millimeter
emission from a forming galaxy occurs before the black hole has grown
large enough to become a quasar. Therefore when most quasars are
shining in the optical they have already exhausted much of their dust
which has been expelled by starburst winds or become locked up in
stars. In this model, the BALs could still be formed at the special
epoch outlined previously, but the dust mass at this epoch may not be
that different to later on when a non-BAL quasar is observed. The main
factor for whether a quasar shows strong sub-millimeter emission or not would
just depend upon how de-coupled the star-formation and AGN epochs are.

\section{Conclusions}

We have conducted the first systematic survey of sub-millimeter
emission from broad absorption line quasars. By comparison with
matched samples of quasars which do not show the BAL phenomenon, we
draw the following conclusions:

\begin{itemize}

\item{The sub-millimeter luminosities of BAL quasars are indistinguishable
from those of quasars without BALs.}

\item{Since sub-millimeter emission is optically-thin and isotropic these
results are consistent with the scenario where all quasars contain
broad absorption line gas, but only a fraction with a particular
orientation actually show BALs in their spectra.} 

\item{The data are inconsistent with a model in which the BAL
phenomenon indicates a special evolutionary stage which co-incides
with a large dust mass in the host galaxy and a high sub-millimeter luminosity.}

\item{Our results provide constraints on alternative models whereby
BALs occur at a special stage in the evolution of quasars.}

\end{itemize}



\acknowledgments

The JCMT is operated by the Joint Astronomy Centre in Hilo, Hawaii on
behalf of the parent organizations Particle Physics and Astronomy
Research Council in the United Kingdom, the National Research Council
of Canada and The Netherlands Organization for Scientific Research.
We are very grateful to the staff at the Joint Astronomy Centre and
Canadian queue observers for their help with the observations. We
thank the anonymous referee for a critical appraisal of the
manuscript. CJW thanks the National Research Council of Canada for
support.

\end{document}